\documentclass[pra,twocolumn,tightenlines,showpacs,nofootinbib]{revtex4}
\usepackage{multirow}
\usepackage{bm,dcolumn,amsmath,graphicx}

\begin{document}

\title{
The magic road to precision}

\author{M. S. Safronova$^{1,2}$}
\author{Z. Zuhrianda$^1$}
\author{U. I. Safronova$^3$}
\author{Charles W. Clark$^2$}

\affiliation{$^1$Department of Physics and Astronomy, University of Delaware, Newark, Delaware 19716}
\affiliation{$^2$Joint Quantum Institute, National Institute of Standards and Technology and the University of Maryland,
Gaithersburg, Maryland, 20899-8410}
\affiliation{$^3$Physics Department, University of Nevada, Reno, Nevada 89557}

\date{\today}

\begin{abstract}
We predict a sequence of magic-zero wavelengths for the Sr excited $5s5p ^3P_0$ state,
and
 provide a general roadmap for extracting transition
matrix elements  using precise frequency measurements.
We demonstrate that such measurements can serve as a best global benchmark of the spectroscopic accuracy that is
required for the  development of high-precision predictive methods.
 These magic-zero wavelengths are also needed for  state-selective atom manipulation for implementation
of  quantum logic operations. We also identify five magic wavelengths of the $5s^2\
^1S_0 - 5s5p\ ^3P_0$ Sr clock transition between 350~nm and 500~nm
 which can also serve as precision benchmarks.

\end{abstract}
\pacs{37.10.Jk, 37.10.De, 32.10.Dk, 31.15.ac}
\maketitle
The drive for increased precision in atomic, molecular, and optical  (AMO)
  measurements has led to tighter tests of fundamental physics \cite{GriSwaLof09,ACME14,PruRamPor15},
  transformational improvements in time and frequency metrology \cite{NicCamHut15,BelHinPhil14}, and suppression of  decoherence in
  quantum information processing \cite{ZhaRobSaf11,GolNorKol15}. Moreover, increases in AMO precision have  resulted in the discovery of new applications such as: laboratory tests of time-variation of fundamental constants \cite{RosHumSch08,GodNisJon14,HunLipTam14}, searches for topological dark matter with atomic clocks \cite{DerPos14} and  magnetometers \cite{PusKimPan13}, probes of gravity and general relativity with atomic interferometry \cite{SchHarAlb14,RosSorCac14,TarMazPol14}, and use of decoherence-free subspaces  for Lorentz symmetry tests with entangled trapped ions \cite{PruRamPor15}.

For many of those applications, accurate knowledge of atomic properties has been  critical for the design and interpretation of experiments, quantifying and reducing uncertainties and decoherence, and development of concepts for next-generation experiments and precision  measurement techniques.
Progress in development of high-precision theory \cite{PorBelDer09,mar-two-09,PorSafKoz12} has yielded  accurate predictions of many needed properties. In turn,  high-precision measurements \cite{WooBwnCho97,BarStaLem08,LudZelCam08,KeeHanWoo11,MidFalLis12,BelSheLem12} have provided experimental benchmarks for refinement and improvement of theory.

Determination of transition matrix elements between excited states is a particulary difficult challenge for both theory and experiment.
 For example,
the accuracy of theory has reached ~0.2\% for low-lying state transitions
of alkali-metal atoms \cite{PorBelDer09} and has attained  1\% to a few percent accuracy for more complicated systems \cite{SafDzuFla14,SafPorCla12,SafKozCla11}.
At present, further progress is hindered by scarcity of high-precision  (better than 1\%) benchmarks of transition amplitudes, which are important in many of the applications cited above.
Heretofore, determination of transition matrix elements between excited states have been based primarily
 upon the measurements of lifetimes and branching ratios, which seems very difficult to push beyond 1\% accuracy, with
 Very few such measurements are available, including the most recent $^3D_1$ lifetime measurement in Sr \cite{NicCamHut15}, where 0.5\% uncertainty was achieved
 for the improved evaluation of the dynamic black-body shift.
  Moreover, such techniques face irreducible difficulties if the relevant branching ratios are small.
  This problem is particularly difficult for transitions involving
   excited states.

In this work, we  provide a roadmap for extracting transition matrix elements using spectroscopic measurements of \textit{frequency}.
We make use of ``magic-zero'' frequencies at which the frequency-dependent polarizability, $\alpha(\omega)$, of a given atomic state vanishes \cite{LebThy07,SafKozCla11}.
These magic-zero frequencies,
which are analogues of interference phenomena encountered in classical coupled LC electrical circuits,
are ubiquitous in accessible regions of the optical spectrum. We demonstrate that such frequency  measurements can serve as a best global benchmark of high-precision atomic theory.

Two magic-zero wavelengths for the ground state of alkali-metal atoms have been recently measured, in Rb  \cite{HerVaiLi12} and K \cite{HolTruHro12}. Thus, there is established experimental methodology  to measure these quantities. The Rb measurement, which is based on finding
the null point of diffraction of ultracold atoms by an optical lattice,
 was used to
determine $5s-6p$ matrix elements as was proposed in \cite{AroSafCla11}.
That determination attained an accuracy of 0.15-0.3\% \cite{HerVaiLi12}.

In this paper, we discuss a systematic approach to the determination of
multiple transition probabilities using magic-zero spectroscopy and we
describe its application to the important test case of the $5s5p ~^3P_0$ state
of Sr. At present, this is one the best available benchmark systems for the following reasons:
   \begin{itemize}
   \item Well-developed experimental techniques are already established for Sr
   due to its prominence in atomic clock development  \cite{NicCamHut15} and studies of many-body effects in degenerate quantum gases \cite{ZhaBisBro14,StePasGri13}.
       \item The  matrix element for the lowest-energy relevant Sr transition, $5s5p ^3P_0$-$5s4d$~$^3D_1$, is known with 0.23\% accuracy from a recent $5s4d~^3D_1$ lifetime measurement \cite{NicCamHut15}.
          This will simplify the extraction of matrix elements from $^3P_0$ to higher excited states.
        \item From a theoretical prospective, Sr is one of the best understood  atoms with more than one valence electron, due to  recent calculations of blackbody radiation (BBR) shift of the $5s^2$~$^1S_0$-$5s5p ~^3P_0$  atomic clock transition \cite{SafPorSaf13}.
   \end{itemize}

 In addition to precision measurement, there is another important application of the magic-zero wavelengths in the alkaline-earth atoms.
The trapping potential in an optical lattice for a given atomic state
 is proportional to its dynamic polarizability $\alpha(\omega)$. For the magic-zero wavelength, $\alpha(\omega)=0$, resulting in
a vanishing ac Stark shift of that  state. Thus, atoms in that state are insensitive to laser light of that frequency.
This enables state-selective atom manipulation for the implementation
of the quantum logic operations \cite{prl17504,GorReyDal09}.
In this work, we locate all of magic-zero wavelengths above 350~nm for both the ground and $^3P_0$ excited
state of Sr.

 Unless stated otherwise, we use the conventional system of atomic units,
a.u., in which $e, m_{\rm e}$, $4\pi \epsilon_0$ and the reduced
Planck constant $\hbar$ have the numerical value 1.
Polarizability in a.u. has the dimension of volume, and its
numerical values presented here are expressed in units of $a^3_0$,
where $a_0\approx0.052918$~nm is the Bohr radius. The atomic units
for $\alpha$ can be converted to SI units via
 $\alpha/h$~[Hz/(V/m)$^2$]=2.48832$\times10^{-8}\alpha$~[a.u.], where
 the conversion coefficient is $4\pi \epsilon_0 a^3_0/h$ and the
 Planck constant $h$ is factored out. Vacuum values are reported for all  wavelengths.
\begin{figure}[tbp]
            \includegraphics[scale=0.32]{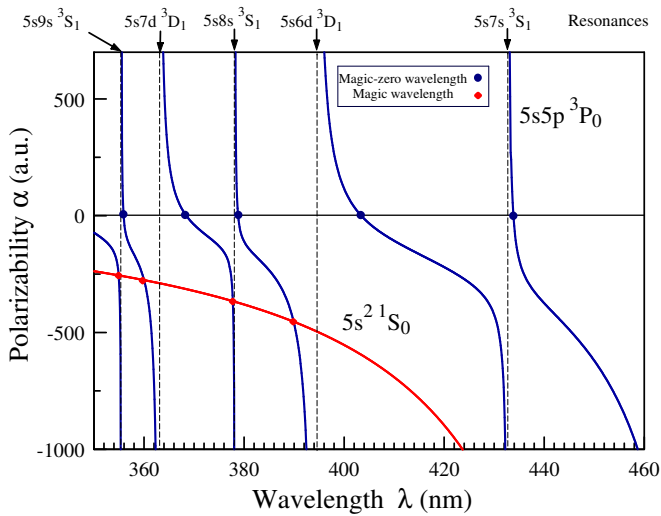}
            \caption{The dynamic polarizabilities of Sr  $5s^2$~$^1S_0$  and $5s5p ^3P_0$ states  in the 350-460~nm wavelength range. }
\label{fig1}
\end{figure}
\begin{figure}[tbp]
            \includegraphics[scale=0.32]{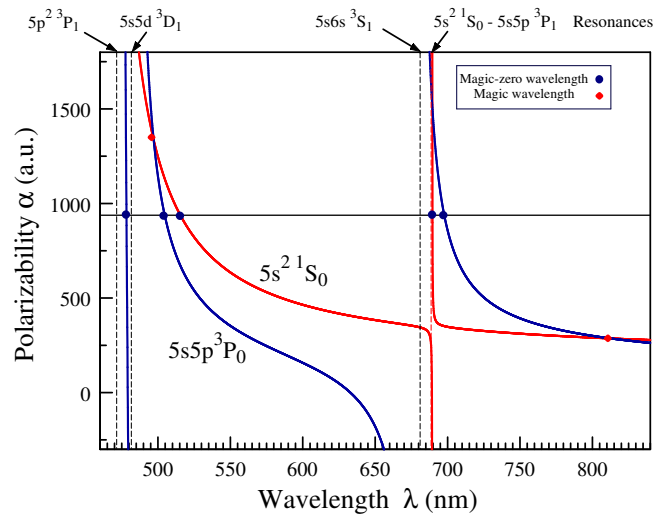}
            \caption{The dynamic polarizabilities of Sr  $5s^2$~$^1S_0$  and $5s5p ^3P_0$ states  in the 460-840~nm wavelength range. }
\label{fig2}
\end{figure}

We now demonstrate how to extract a set of matrix elements from
 a sequence of magic-zero wavelengths for Sr  $5s5p ~^3P_0$ state.
 Consider the standard expression for a frequency-dependent polarizability of a state $v$ in terms of the
 sum over all other atomic states $k$
 ~\cite{MitSafCla10}:
\begin{equation}
\alpha _{0}^{v}(\omega
)=\frac{2}{3(2J+1)}\sum_{k}\frac{{\left\langle
k\left\| D\right\| v\right\rangle }^{2}(E_{k}-E_{v})}{(E_{k}-E_{v})^{2}-%
\omega ^{2}}.  \label{eq-pol}
\end{equation}
Here, $J$ is the total angular moment of the state $v$ and
${\left\langle k\left\|D\right\|v\right\rangle}$ are the reduced
electric-dipole matrix elements which are the subject of the present work. In these equations, $\omega$ is
assumed to be at least several line widths off resonance with the
corresponding transitions. Linear polarization is assumed in all
calculation.

It is self-evident from Eq.~(\ref{eq-pol}) that the contribution of state $k$ to the sum changes sign as
$\omega$ crosses the value $E_{k}-E_{v}$.
 \textit{Magic-zero wavelengths} arise due to the cancellation of the contribution from a given resonant state $k$ with the contributions from
 all of the other resonant  states.
 Thus,
there is a magic-zero wavelength between each pair of adjacent resonances, as can be seen in Figs.\ref{fig1} and \ref{fig2}.
The actual location of each magic zero wavelength depends upon the distribution of electric-dipole matrix elements
${\left\langle k\left\|D\right\|v\right\rangle}$.
 Different matrix elements
 will be important for the determination of each magic zero-wavelength. Thus,
  measurement of a series of  magic-zero wavelengths will enable one to
extract the entire recommended set of the matrix elements  for transitions to the $^3P_0$ state.
Furthermore, we can obtain complementary information from
\textit{magic wavelengths} of the Sr $5s^2\
^1S_0 - 5s5p\ ^3P_0$ clock transition, at which
 Stark shift of the clock transition vanishes.
  With such  high-precision benchmarks established in one system to test a new theoretical approach,
   the theory can be applied to a large number of other systems
   where no experimental data are available.

Sr has two valence electrons outside a closed Kr-like atomic core. The main challenge in the
theoretical treatment of systems with two or  more valence electrons is the accurate treatment of both core-valence
correlations and strong valence-valence correlations.
We use the hybrid approach
introduced in \cite{mar-two-09} that combines configuration interaction (CI) and  all-order linearized
coupled-cluster methods. The core-valence (and core-core) correlations are treated by the coupled-cluster all-order method, which is used to
construct the effective Hamiltonian.
 Then, this effective Hamiltonian is used in the configuration-interaction part of the method that treats the
valence-valence correlations. Thus, all of the correlation correction to the wave functions is
treated at the all-order level.
Next, the resulting wave functions are used to evaluate matrix elements of various one-body operators, such as electric and magnetic mulitpole, magnetic and quadrupole hyperfine, and various P-odd and T-odd interactions.  These matrix elements are also then used for
evaluation of polarizabilities, P-odd and T-odd amplitudes, and long-range interaction coefficients $C_6$ and $C_8$.
This approach is generally applicable for systems with several valence electrons and has a wide range of applications \cite{SafDzuFla14,SafPorCla12,SafKozCla11}.

To determine the magic-zero wavelengths for the Sr $5s^2$~$^1S_0$ ground and $5s5p ^3P_0$ excited states, we need to
calculate their  frequency dependent polarizabilities for a wide range of frequencies.
 The valence part of the polarizability is determined  by
solving the inhomogeneous equation of perturbation theory in the
valence space, which is approximated as
\begin{equation}
(E_v - H_{\textrm{eff}})|\Psi(v,M^{\prime})\rangle = D_{\mathrm{eff},q} |\Psi_0(v,J,M)\rangle
\label{eq1}
\end{equation}
for a state  $v$ with the total angular momentum $J$ and projection
$M$ \cite{kozlov99a}.  While the $H_{\textrm{eff}}$ includes the all-order corrections as described above,
the effective dipole operator
$D_{\textrm{eff}}$ only includes random phase approximation (RPA) corrections at the present time.

 A few other small corrections can be calculated with second-order  many-body theory.
   These corrections to dipole matrix elements contribute significantly in the $5s5p$~$^3P_0$ BBR shift calculation of Sr~\cite{SafPorSaf13}.  Their omission in earlier work \cite{PorDer06} resulted in significant difference of the clock dc Stark shift with experiment ~\cite{MidFalLis12}. The corrections are very significant for the hyperfine constants \cite{kozlov99}.

The present bottleneck in the accuracy of this approach can be summarized as follows. The treatment of corrections to the matrix elements of the one-body electric-dipole operator $D_{\textrm{eff}}$, and all of the other one-body operators mentioned above is limited to RPA and second-order many-body perturbation theory.
We are currently developing a full all-order treatment of corrections to one-body operators, however there are essentially no experimental benchmarks that we can use to establish how well this approach will work. The measurements  that we propose in this work will remedy this outstanding problem.  We note that  the method is non-specific for the particular type of the one-body operator and the advancement in the treatment of the  dipole operator will also provide improvement for the other operators.

While we  calculate the polarizabilities by solving the inhomogeneous equation (\ref{eq1}), which accounts for the contribution from all bound  and continuum states, we extract several dominant contributions from the low-lying bound states using the sum-over-states  formula (\ref{eq-pol}).
To improve the accuracy of the calculations, we replace a few dominant terms in our polarizability calculations by their ``recommended'' values, which
contain experimental energies and recommended matrix elements from Ref.~\cite{SafPorSaf13}, where available.
To obtain recommended values for the $^3P_0$ state, the energies  of the ten lowest transitions are replaced by the experimental values \cite{nist-web}, five of the transition matrix elements  are replaced by the recommended values from Ref.~\cite{SafPorSaf13}, and $5s5p~^3P_0-5s4d~^3D_1$ matrix elements is taken from \cite{NicCamHut15}.
For the $^1S_0$ state, the energies  of the four lowest transitions  and the  $5s^2~^1S_0-5s5p~^1P_1$ matrix element is replaced by the experimental values \cite{nist-web,YasKisTak06}.
Both \textit{ab initio} and recommended values are listed in Table~\ref{tab1}.

\begin{table} \caption{\label{tab1}
Resonance wavelengths $\lambda$ and
  reduced dipole matrix elements $D$ in Sr. Vacuum wavelength values are given in nm. The recommended set of the matrix elements in a.u. is from   \cite{SafPorSaf13}.
  $^a$Ref.~\cite{NicCamHut15}. }
\begin{ruledtabular}
\begin{tabular}{lllllll}
\multicolumn{3}{c}{Transition}&
\multicolumn{2}{c}{Wavelength} &
\multicolumn{2}{c}{Matrix elements $D$}\\
\multicolumn{3}{c}{}&
\multicolumn{1}{c}{CI+all}&
\multicolumn{1}{c}{Expt.}&
\multicolumn{1}{c}{CI+all}&
\multicolumn{1}{r}{Recomm.}
\\
\hline
$5s5p\ ^3P_0  $&-&$ 5s4d\ ^3D_1$&     2642.4  &    2603.1 &    2.714  &   2.6707(62)$^a$   \\
$5s5p\ ^3P_0  $&-&$ 5s6s\ ^3S_1$&     682.0   &    679.3  &    1.972  &   1.962(10)    \\
$5s5p\ ^3P_0  $&-&$ 5s5d\ ^3D_1$&     484.2   &    483.3  &    2.458  &   2.450(24)    \\
$5s5p\ ^3P_0  $&-&$ 5p^2 \ ^3P_1$&     471.5   &    474.3  &    2.627  &   2.605(26)    \\
$5s5p\ ^3P_0  $&-&$ 5s7s\ ^3S_1$&     433.3   &    432.8  &    0.522  &   0.516(8)    \\
$5s5p\ ^3P_0  $&-&$ 5s6d\ ^3D_1$&     394.3   &    394.2  &    1.175  &   1.161(17)    \\
$5s5p\ ^3P_0  $&-&$ 5s8s\ ^3S_1$&     377.1   &    378.2  &    0.302  &            \\
$5s5p\ ^3P_0  $&-&$ 5s7d\ ^3D_1$&     361.4   &    363.0  &    0.822  &            \\
$5s5p\ ^3P_0  $&-&$ 5s9s\ ^3S_1$&     348.4   &    355.4  &    0.270  &            \\
$5s5p\ ^3P_0  $&-&$ 5s8d\ ^3D_1$&     336.7   &    334.8  &    0.820  &
\end{tabular}
\end{ruledtabular}
\end{table}

\begin{table} \caption{\label{tab2}
Magic-zero $\lambda_{\rm zero}$ and magic $\lambda_{\rm magic}$    wavelengths.
 See text for the explanation of the recommended value calculations.  }
\begin{ruledtabular}
\begin{tabular}{llllll}
\multicolumn{2}{c}{$\lambda_{\rm zero}$}&
\multicolumn{2}{c}{$\lambda_{\rm magic}$}\\
\multicolumn{1}{c}{CI+all-order}&
\multicolumn{1}{c}{Recomm.}&
\multicolumn{1}{c}{CI+all-order}&
\multicolumn{1}{c}{Recomm.}
\\ \hline
\multicolumn{2}{c}{$5s5p~^3P_0$} &\multicolumn{2}{c}{$5s^2\ ^1S_0-5s5p~^3P_0$}\\
       & 355.92        &          &  354.9         \\
367.0  & 368.45   &   358.5  &  360.0 \\
377.8  & 378.81    &   376.8  &  377.75 \\
403.35 & 403.428    &   390.1  &  389.9 \\
434.35 & 433.85    &           & 497.0        \\
478.35 & 479.126    &           &         \\
634.7  & 632.83    &           &         \\
1672.9 & 1666.6   &           &         \\
 \multicolumn{2}{c}{$5s^2\ ^1S_0$}&&          \\
679.55& 689.20&  &        \\
\end{tabular}
\end{ruledtabular}
\end{table}

\begin{table*} \caption{\label{tab3}
 Breakdown by transition of the contributions (in a.u.) to  dynamic polarizability of $5s5p~^3P_0$ state, at the eight magic-zero wavelengths indicated.
 The first ten rows give the  contributions from the transitions indicated, and all other contribution are grouped together in row ``Other''. The chain of dominant contributions relevant to the extraction of matrix elements (see text for a discussion) is highlighted in bold.}
\begin{ruledtabular}
\begin{tabular}{lccrrrrrrrr}
\multicolumn{3}{c}{Contribution}&
 \multicolumn{1}{c}{1666.6~nm}&
    \multicolumn{1}{c}{632.84~nm}&
 \multicolumn{1}{c}{479.127~nm}&
 \multicolumn{1}{c}{433.85~nm}&
\multicolumn{1}{c}{403.429~nm}&
\multicolumn{1}{c}{378.81~nm}&
 \multicolumn{1}{c}{368.45~nm} &
  \multicolumn{1}{c}{355.92~nm}  \\
\hline
$5s5p\ ^3P_0  $&-&$ 5s4d\ ^3D_1$&  \textbf{-188.7} &   -17.1  &     -9.5  &      -7.8 &      -6.7  &    -5.9  &    -5.6 &    -5.2   \\
$5s5p\ ^3P_0  $&-&$ 5s6s\ ^3S_1$&   \textbf{45.9} &   \textbf{-251.4}  &    -37.9  &     -26.4 &     -20.8  &   -17.3  &   -15.9 &   -14.5   \\
$5s5p\ ^3P_0  $&-&$ 5s5d\ ^3D_1$&    \textbf{ 46.3} &    \textbf{101.9}  &   \textbf{ -2404} &     -176.0&     -97.5  &   -67.6  &   -58.9 &   -50.3   \\
$5s5p\ ^3P_0  $&-&$ 5p^2\ ^3P_1$&    \textbf{ 51.2} &    \textbf{107.5}  &     \textbf{2361} &     -241.2&    -123.2  &   -82.9  &   -71.7 &   -60.7   \\
$5s5p\ ^3P_0  $&-&$ 5s7s\ ^3S_1$&     1.8 &    3.2   &      9.2  &      \textbf{337.9} &     -11.2  &    -5.5  &    -4.4 &    -3.5   \\
$5s5p\ ^3P_0  $&-&$ 5s6d\ ^3D_1$&     8.2 &   12.7   &     24.1  &      44.6 &      \textbf{171.7}  &   -93.8  &   -53.8 &   -34.3   \\
$5s5p\ ^3P_0  $&-&$ 5s8s\ ^3S_1$&     0.5 &   0.8    &      1.3  &       2.1 &       4.2  &    \textbf{147.2}  &    -9.5 &    -3.9   \\
$5s5p\ ^3P_0  $&-&$ 5s7d\ ^3D_1$&     3.8 &   5.4    &      8.4  &      12.0 &      18.9  &    44.0  &   \textbf{ 122.7 }&   -89.1   \\
$5s5p\ ^3P_0  $&-&$ 5s9s\ ^3S_1$&     0.4 &   0.6    &      0.8  &       1.2 &       1.7  &     3.2  &     5.5 &    \textbf{142.7 }  \\
$5s5p\ ^3P_0  $&-&$ 5s8d\ ^3D_1$&     3.4 &   4.6    &      6.4  &       8.1 &      10.6  &    15.1  &    18.9 &    28.6   \\
Other           & &             &    27.1 &   32.0   &     39.2  &      44.8 &      52.3  &    64.0  &    72.7 &    90.6    \\
Total           & &             &     0.0 &   0.0    &     -0.3  &      -0.7 &      0.0   &     0.4  &     0.0 &     0.4 \\ \hline
\multicolumn{3}{l}{Uncertainty in $\alpha$}
&                                11.6      &  8.7 &       10.5  &   6.6     &    12       & 67      &      4.3 &   2.2 \\
\multicolumn{3}{l}{Uncertainty in $\lambda_{zero}$}
&                                0.1~nm   &0.2~nm  & 0.05~nm    & 0.25~nm&     0.05~nm   & 0.1~nm& 0.6~nm& 6~nm\\
\end{tabular}
\end{ruledtabular}
\end{table*}

The dynamical polarizabilities of Sr $5s^2$~$^1S_0$ ground and $5s5p ^3P_0$ excited states are plotted in Figs.~\ref{fig1} and \ref{fig2}.
For clarity, we separate these into two wavelength region; 350-460~nm and 460-840~nm. All resonances from Table~\ref{tab1} are labeled on the top frame of the figures.
 For all of  the $^3P_0$ resonances, we only write the upper transition states and omit
$5s5p ~^3P_0$ designations. The magic-zero wavelengths are determined by locating points where either $^3P_0$ or $^1S_0$ polarizabilities vanish. We also show magic wavelengths, where the $^3P_0$ and $^1S_0$ polarizabilities are identical. The 813.4~nm magic wavelength used for the Sr lattice clock  \cite{LudZelCam08} is shown near the right edge of Fig.~\ref{fig2}.
We note that our graphs show 5 other Sr magic wavelengths.
The values of both magic-zero and magic wavelengths are summarized in Table~\ref{tab2}.  Both \textit{ab initio} and recommended values are listed.
Our value of the $5s^2$~$^1S_0$ magic-zero wavelength agrees with the results of \cite{prl17504,pra022511}. Ref.\cite{prl17504} estimate of the
$^3P_0$ magic-zero wavelength at 627~nm is 5~nm away from our 632.8(2)~nm value.

The contributions from the 10 lowest resonant states to the $5s5p~^3P_0$ polarizabilities  at the magic-zero wavelengths
are given in the first 10 rows of Table~\ref{tab3}. The exact values of those wavelengths are given as the labels of the eight columns.  All of the other contributions  are grouped together in the row ``Other''. This contribution smoothly varies with the wavelength and is substantially smaller than the dominant contributions for most magic-zero wavelengths. The sum of all contributions is zero within the numerical accuracy.

Table~\ref{tab3} demonstrates the use of
the magic-zero wavelength measurements to establish experimental benchmarks for the transitions noted in bold font. First $5s5p\ ^3P_0  - 5s4d\ ^3D_1$ matrix element is already known from a precision lifetime measurement \cite{NicCamHut15}, so contributions in the first row are known with 0.5\% precision. The matrix element of the second transition $5s5p\ ^3P_0  - 5s6s\ ^3S_1$ is constrained to 0.5\% by the position of the 813.4~nm magic wavelength \cite{SafPorSaf13}; thus the values in the second row are known to about 1\%.

We propose two methods for the extraction of the remaining matrix elements. First, a global fit of all measured magic-zero wavelength can be done, varying the dominant contributions to best match the experimental values of wavelength.
Second, a simpler procedure can be used to extract the matrix elements sequentially by
determination of dominant contributions, as follows

\noindent \textbf{1.} The dominant contributions  of the third and fourth transitions in Table~\ref{tab3}, associated with the  $5s5d\ ^3D_1$ and $5p^2\ ^3P_1$ states respectively, can be extracted from measurements of the first three  magic-zero wavelengths. If all three are available, the uncertainty in the ``Other'' contribution can be established as well.
The 479~nm wavelength is particulary useful, due to very large contributions of the third and fourth transitions. The correlation corrections
are expected to be different for the transition to the $5p^2\ ^3P_1$ state in comparison to all other states, due to its different electron configuration
(see Table~II of Ref.~\cite{SafPorSaf13}). Therefore, we recommend the precision measurement of 479~nm wavelength as a first priority.

\noindent \textbf{2.} With first four dominant contributions in hand, the fifth and sixth can be obtained
independently from the 434~nm and 403~nm magic-zero wavelengths since only one of these transitions contributes significantly
to its respective magic-zero wavelength. This happens because of the larger values of the $^3D_1$ matrix elements in comparison with those of $^3S_1$ for the higher states (see Table~\ref{tab1}).

\noindent \textbf{3.} Then, dominant contributions for the $5s8s ^3S_1,$ $5s7d ^3D_1$, and $5s9s ^3S_1$ transitions can be obtained  if the last three wavelengths are known.


\noindent The magic wavelengths listed in Table~\ref{tab2} can be used as  additional benchmarks in a similar way.

The uncertainties in the polarizability values and resulting estimated uncertainties in the magic-zero wavelengths are listed in the last two rows of Table~\ref{tab3}. The polarizability uncertainties are obtained by adding the estimated uncertainties
in the each of ten contributions and uncertainties of the ``Other'' term in quadrature.
The relative uncertainties in the polarizability contributions
are  twice the estimated relative    uncertainties of the corresponding matrix elements listed in Table~\ref{tab1}. The uncertainties in the last four matrix elements were taken to be 3\% and the uncertainty in the ``Other'' contribution was estimated at 5\%.

In summary, we have predicted the values of nine magic-zero and five magic wavelengths for Sr. We demonstrated how measurements of these quantities
may serve as a sensitive  experimental benchmark for further development of higher-precision first-principles theory.


 This research was
performed under the sponsorship of the US Department of Commerce,
National Institute of Standards and Technology, and was supported by the National
Science Foundation under Physics Frontiers Center Grant PHY-0822671.


\end{document}